\documentclass[10pt,notitlepage,showpacs,pra,twocolumn]{revtex4-1}
\usepackage{ulem}
\usepackage{amsfonts}
\usepackage{amsmath}
\usepackage{amssymb}
\usepackage{graphicx}
\usepackage{float}
\usepackage{color}
\usepackage[toc,page,header]{appendix}

\begin{document}

\title{Properties of collective Rabi oscillations with two Rydberg atoms}
\author{Dandan Ma, Keye Zhang$^{*}$, Jing Qian$^{\dagger}$ }
\affiliation{Department of Physics, School of Physics and Material Science, East China
Normal University, Shanghai 200062, People's Republic of China}

\begin{abstract}
Motivated by experimental advances [e.g. A. Ga{\"e}tan {\it et.al.} Nat. Phys. 5 115 (2009)] that the collective excitation of two Rydberg atoms was observed, we provide an elaborate theoretical study for the dynamical behavior of two-atom Rabi oscillations. In the large-intermediate-detuning case, the two-photon Rabi oscillation is found to be significantly affected by the strength of the interatomic van der Waals interaction. With a careful comparison of the exact numbers and values of the oscillation frequency, we propose a new way to determine the strength of excitation blockade, well agreeing with the previous universal criterion for full, partial and none blockade regions. In the small-intermediate-detuning case we find a blockade-like effect, but the collective enhancement factor is smaller than $\sqrt{2}$ due to the quantum interference of double optical transitions involving the intermediate state. Moreover, a fast two-photon Rabi oscillation in $ns$ timescale is manifested by employing intense lasers with an intensity of $\sim$MW/cm$^2$, offering a possibility of ultrafast control of quantum dynamics with Rydberg atoms.
\end{abstract}
\email{jqian1982@gmail.com}
\maketitle
\preprint{}

\section{Introduction}
Observing Rabi oscillations of high contrast is a fundamental step for realizing efficient population transfer between quantum states, being essential to the adiabatic population transfer by stimulated Raman adiabatic passage \cite{Vitanov17}, and to the control of coherent quantum systems such as quantum dots \cite{Petta05,Koppens06,Pellegrino18}, solid-state systems \cite{Greenland10,Lillie17,Zhou17}, and nuclear ensembles \cite{Morley13,Haber17,Sigillito17}.

Systems of cold Rydberg atoms provide an excellent platform for studying collective many-body phenomena owing to their exotic properties \cite{Gallagher94,Low12}. When two Rydberg atoms are close to each other, their simultaneous excitation driven by a same laser pulse may be forbidden, meanwhile the single Rydberg excitation of  one atom is coherently enhanced. This phenomenon is known as ``excitation blockade" \cite{Lukin01,Comparat10}, which has been employed to produce entangled state \cite{Gaetan09,Wilk10} and quantum CNOT gate \cite{Urban09,Isenhower10}. An important extension for studying excitation blockade is to adopt a blockaded ensemble {\it i.e.}``superatom" to explore the many-body effect \cite{Garttner14,Weber15,Ebert15,Mandoki17}. The coherent ground-Rydberg Rabi oscillation, as a main representation for collective dynamical behavior of Rydberg atoms driven by the radiation fields \cite{Gentile89}, is crucial for the manipulation of Rydberg-Rydberg interactions \cite{Johnson08,Reinhard08, Labuhn16}, and for the applications of Rydberg atoms in quantum information processing \cite{Deiglmayr06,Patton13,Saffman10}.

Recent experiments observed that the optically driven Rabi oscillation of two or more Rydberg atoms have a high contrast, which is perfectly consistent with the theoretical predictions. Due to the excitation blockade, a two-level Rydberg atomic ensemble can support a coherent one-photon Rabi oscillation with a frequency of $\sqrt{N}\Omega$ ($\Omega$ is the ground-Rydberg Rabi frequency for a single atom, $N$ is the atomic number) \cite{Stanojevic09,Pritchard10,Dudin12}. 
The extension to the multiphoton regime that enables a robust two-photon Rabi oscillation driving the population transfer on an extremely short timescale, was also experimentally investigated \cite{Fushitani16}. And very recently, experimentalists noted the problem that the influences of spontaneous emission and ac Stark shifts from the intermediate state can be eliminated by a single-photon approach \cite{Hankin14}, and analyzed various imperfections for damping and finite contrast in coherent optical excitation of Rydberg state \cite{Leseleuc18}.

In this article, we study the coherent ground-Rydberg Rabi oscillations in two collectively-excited atoms which are of three-level structure in a more detailed way. Compared to a recent relevant work \cite{Beguin13} where authors experimentally exploited the properties of Rabi oscillation by changing the strength of excitation blockade and a self-consistent theoretical model based on two two-level atoms was given for the results, here, we employ the model of two three-level atoms to simulate the collective excitations, perfectly reproducing their experimental observations. Furthermore, the properties of coherent Rabi oscillation in both large- and small-intermediate-detuning cases are clearly demonstrated. Specially, we first analyze a single-atom case where the single-atom Rabi oscillation frequency is found to decrease with the increase of the intermediate detuning.
While turning to the two three-level-atom case, for a large intermediate detuning, an effective two-level system is supported, giving rise to a collective Rabi oscillation with an enhanced frequency. Our numerical simulations show that the strength of excitation blockade can be distinguished by the numbers and values of Rabi oscillation frequencies, which well agrees with the universal criterion by comparing the relative strength of the interatomic van der Waals(vdWs) interaction and the effective off-resonant Rabi frequency. For the case of a small intermediate detuning, we present a fast two-photon Rabi oscillation in $n$s timescale via intense laser fields under realistic experimental parameters. Also an excitation blockade-like effect can be obtained when the strength of the vdWs interaction is dominant. However, due to the quantum interference between two optical transition paths involving the intermediate state, the enhancement factor for the collective Rabi frequency is smaller. Such a fast Rabi oscillation is significant for achieving coherent manipulation of quantum states in an atomic system with large spontaneous decays.

\section{Model and single-atom case} 

\begin{figure}
\centering
\includegraphics[width=3.4in,height=2.9in]{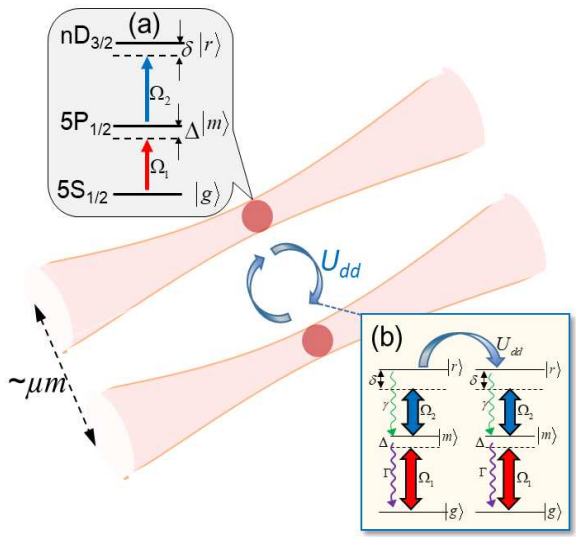}
\caption{(color online) Proposal for the experimental implementation. Two isolated atoms are loaded in tightly focused dipole traps, {\it i.e.}, ``optical tweezers" separated by a distance of $\sim\mu$m. The vdWs interaction $U_{dd}$ causes an energy shift to one of atomic state $\left\vert r\right\rangle$, obstructing the simultaneous Rydberg excitation of two atoms.
Our detection is mainly carried out on the time-dependent population dynamics of singly-excited and doubly-excited Rydberg states. Insets (a-b) represent the atom-field interactions and the vdWs interactions in the single- and two-atom pictures. Relevant parameters are described in the main text.} 
\label{model}
\end{figure}

We consider that the energy levels of a single three-level atom $^{87}$Rb consists of the ground state $\left\vert g\right\rangle=\left\vert 5S_{1/2}\right\rangle$, the intermediate state $\left\vert m\right\rangle=\left\vert 5P_{1/2}\right\rangle$, and the Rydberg state $\left\vert r\right\rangle=\left\vert nD\right\rangle$ (or $\left\vert nS\right\rangle$) \cite{Miroshnychenko10}, see the inset (a) of Fig. \ref{model}. The excitation to $\left\vert r\right\rangle$ is accomplished by a two-step optical excitation with the pump and coupling laser Rabi frequencies $\Omega_{1}$ and $\Omega_{2}$, detuned by $\Delta$ and $\delta$ from $\left\vert m\right\rangle$ and $\left\vert r\right\rangle$, respectively. Due to the fact that a direct one-photon Rydberg excitation requires a very powerful UV laser and is limited by its transition selection rules when $\left\vert r\right\rangle=\left\vert nP\right\rangle $ \cite{Wang17}, most current experiments adopted a two-photon scheme for its suitable energy level and easy optical control. A similar system based on two Y-typed four-level atoms was developed to achieve a reversible switching of population on two different Rydberg states with very high fidelity \cite{Jing17}.

We begin with the study of a single atom [see inset (a) of Fig. \ref{model}] where numerical results are obtained by simulating the master equation (ME): $\dot{\rho}=-i[\mathcal{H}_0 ,\rho] + \mathcal{L}_0$, with the one-atom Hamiltonian $\mathcal{H}_0$ and the Lindblad operator $\mathcal{L}_0$ taking the form of ($\hbar = 1$)
\begin{eqnarray}
\mathcal{H}_0 &=&\left( 
\begin{array}{ccc}
0 & \frac{\Omega_1}{2} & 0  \\ 
 \frac{\Omega_1}{2} & -\Delta & \frac{\Omega_2}{2}  \\ 
 0 &  \frac{\Omega_2}{2} &  -\delta   %
\end{array}%
\right)  \label{Ham} \\
\mathcal{L}_0 &=&\left( 
\begin{array}{ccc}
\Gamma\rho_{mm} & -\frac{\Gamma}{2}\rho_{gm} & -\frac{\gamma}{2}\rho_{gr}  \\ 
-\frac{\Gamma}{2}\rho_{mg}& \gamma\rho_{rr}-\Gamma\rho_{mm} & -\frac{\Gamma+\gamma}{2}\rho_{mr}  \\ 
  -\frac{\gamma}{2}\rho_{rg} &  -\frac{\Gamma+\gamma}{2}\rho_{rm} &  -\gamma\rho_{rr}   %
\end{array}%
\right)  \label{Lind}
\end{eqnarray}
where $\Gamma$ and $\gamma$ are the incoherent spontaneous emission rates for two optical transitions $\left\vert m\right\rangle\to\left\vert g\right\rangle$ and $\left\vert r\right\rangle\to\left\vert g\right\rangle$. $\rho_{ii}$ represents the population of $\left\vert i\right\rangle$ and $\rho_{ij}(i\neq j)$ the interstate coherence.

Different from the previous researches of stimulated adiabatic transfer to high Rydberg states \cite{Cubel05} as well as spatial correlations between atoms \cite{Stanojevic09} where Gaussian or square-shaped pulses are applied, here we adopt continuous laser driving and set $\delta=0$ to preserve the coherence between $\left\vert g\right\rangle$ and $\left\vert r\right\rangle$. Note that, if the two-photon excitation is far off-resonant from $\left\vert m\right\rangle$ but resonant to $\left\vert r\right\rangle$, the system could reduce to an effective two-level scheme, in which $\left\vert g\right\rangle$ and $\left\vert r\right\rangle$ are coupled by an effective off-resonant Rabi frequency $\Omega_{eff}$ with detuning $\delta_{eff}$, expressed as \cite{Browaeys16}
\begin{equation}
\Omega_{eff} = \frac{\Omega_1\Omega_2}{2|\Delta|},\delta_{eff} = \delta-\frac{1}{4|\Delta|}(\Omega_1^2-\Omega_2^2),
\label{effect}
\end{equation}
This is a good approximation for studying direct coherent oscillation dynamics between $\left\vert g\right\rangle$ and $\left\vert r\right\rangle$ \cite{Lamour08}.


In the simulations we scale all frequencies (time) by $\Omega_1$ ($\Omega_1^{-1}$) and define the ratio between pump and coupling fields as $\chi=\Omega_1/\Omega_2$. We assume the decay rate $\gamma$ is typically smaller than $\Gamma$ by three orders of magnitude. The oscillating frequency of $\rho_{rr}(t)$ denoted by $\omega_{1,r}$ (the subscript ``r" means singly-excited Rydberg state and ``1" means the one-atom case) is our main observable, which can be extracted from the Fourier-transformed function $f_{r}(\omega_{1,r})$ with respect to $\rho_{rr}(t)$. In general, $f_{r}(\omega_{1,r})$ is continuous and of single-peak structure in the frequency domain. Especially, when the coherent population oscillations have several compatible frequencies, $f_{r}(\omega_{1,r})$ is expected to be a multiple-peak structure in which $\omega_{1,r}^{pk}$ records the dominant frequency at the maximal peak of $f_{r}(\omega_{1,r})$. Otherwise, if $f_r(\omega_{1,r}^{pk})$ is at least one order of magnitude larger than other sub-peak amplitudes $f_r(\omega_{1,r}^{spk})$ ($\omega_{1,r}^{spk}$ is the sub-leading frequency), $f_{r}(\omega_{1,r})$ is expected to be a single-peak function, representing regular Rabi oscillations with single frequency.

\begin{figure}
\centering
\includegraphics[width=3.4in,height=2.6in]{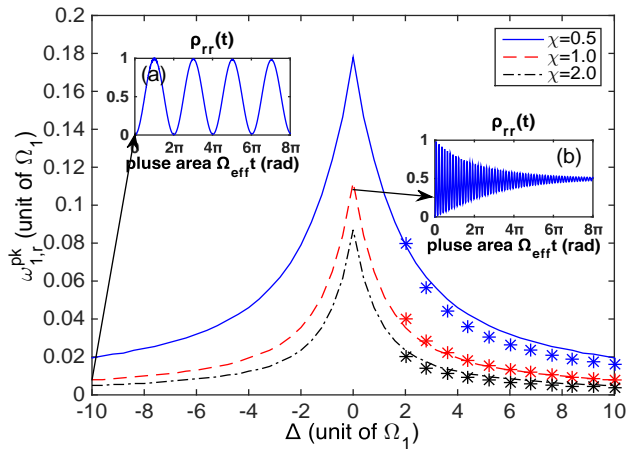}
\caption{(color online). The frequency values $\omega_{1,r}^{pk}$ of the single-atom Rabi oscillation versus $\Delta$ for $\chi=\Omega_1/\Omega_2=0.5$ (blue-solid), 1.0 (red-dashed), 2.0 (black-dot-dashed). Accordingly, the analytical expression $\Omega_{eff}/2\pi$ is plotted by stars with same colors. Dynamical behaviors of two-photon Rabi oscillations in time domain are shown in insets (a) $\Delta/\Omega_1=-10$, $\chi=1.0$ and (b) $\Delta=0$, $\chi=1.0$. The total evolution time is 8 $\mu$s. The decay rates are $\Gamma/\Omega_1=0.5$ and $\gamma/\Omega_1=0.0005$.}
\label{singleatom}
\end{figure}

In Fig. \ref{singleatom}, we show the value of $\omega_{1,r}^{pk}$ by simulating the two-photon excitation process with a tunable detuning $\Delta\in[-10\Omega_1,10\Omega_1]$. 
A clear single-peak frequency function $f_{r}(\omega_{1,r})$ is manifested by the numerical results, with the peak frequency $\omega_{1,r}^{pk}$ varying as a monotone decay with $|\Delta|$. For a large $|\Delta|$, $\omega_{1,r}^{pk}$ approaches a fixed value $\Omega_{eff}/2\pi$ as plotted by stars, coinciding with the two-level approximation at off-resonance cases. As a result, an enhanced correlation between $\left\vert g\right\rangle$ and $\left\vert r\right\rangle$ is robustly established, accompanied by a negligible loss from the intermediate state. Moreover, a high-contrast Rabi oscillation can be achieved with a negligible population damping during the time of pulse area $\Omega_{eff}t$($<8 \mu$s). For example, see inset (a)  it is clear that a non-damping regular Rabi oscillation appears at $\Delta/\Omega_1=-10$ and $\chi=1$.

On the contrary, if $\Delta\to0$, a resonant two-photon process will suffer from a large decoherence due to the effect of the intermediate state, leading to a fast-damped high-frequency Rabi oscillation towards the steady state \cite{Kittelmann96}, see the inset (b) of Fig. \ref{singleatom}. To this end, a robust ultrafast Rabi oscillation on an unprecedented timescale has been achieved in this regime utilizing a strong-field Freeman resonance against the intermediate decay \cite{Fushitani16} or in hot atomic vapor cells via a bandwidth-limited pulse \cite{Huber11}, which offers possibilities for ultrafast quantum state manipulation in Rydberg systems \cite{Li16}. Fig. \ref{singleatom} represents that $\omega_{1,r}^{pk}$ is increased by at least one order of magnitude when $|\Delta|$ is adjusted from far off-resonance ($|\Delta|=10\Omega_1$) to resonance ($\Delta=0$). The increase of $\chi$ leads to a reduction in the frequency, which allows an adiabatic population evolution to the higher states with fewer oscillating circles \cite{Bergmann98}.

\section{Two collectively-excited atoms}

When two trapped atoms occupy the same Rydberg state $|r\rangle$, they will interact via a vdWs potential $U_{dd}=C_6/R^6$ with $R$ the interatomic distance and $C_6$ the corresponding vdWs coefficient. However, when atoms are on different Rydberg states, a resonant dipole-dipole interaction is observed instead \cite{Vogt06}, whose strength is scaled by $R^{-3}$ and can be controlled by external electric fields. Here we only consider the vdWs interaction [inset (b) of Fig. \ref{model}], the Hamiltonian for two interacting atoms is $\mathcal{H}_{I}=\mathcal{H}_{0}\otimes \mathcal{I}+\mathcal{I}\otimes 
\mathcal{H}_{0}+U_{dd}\left\vert rr\right\rangle \left\langle rr\right\vert$ with $ \mathcal{I}$ an identity $3\times 3$ matrix, the subscript $I$ for interaction and $\mathcal{H}_{0}$ given in Eq.(\ref{Ham}). To explore the properties of collective Rabi oscillation between two atoms, we numerically solve the ME in the basis of full vectors, which is $\dot{\rho}_{I} = -i[\mathcal{H}_I,\rho_{I}]+\sum_{j=1,2}\mathcal{L}_{j}[\rho_{I}]$ replacing $\rho$ and $\mathcal{L}_{0}$ by the two-atom density matrix $\rho_{I}$ and dissipative operators $\mathcal{L}_{1,2}$.
The observables we measured in the time domain are the single and double Rydberg populations, denoted by
\begin{eqnarray}
P_{r}(t)&=&\rho_{gr,gr}+\rho_{rg,rg}+\rho_{mr,mr}+\rho_{rm,rm}, \\
P_{rr}(t)&=&\rho_{rr,rr},
\end{eqnarray}
As turning to the frequency domain, $F_{r}(\omega_{2,r})$ and $F_{rr}(\omega_{2,rr})$ are the Fourier transform functions of $P_{r}(t)$ and $P_{rr}(t)$, storing the information of numbers and values of the frequencies. Accordingly, $\omega_{2,r}^{pk}$ and $\omega_{2,rr}^{pk}$ represent the peak frequencies in the two-atom case. Note that $\mathcal{L}_{1}=\mathcal{L}_{2}=\mathcal{L}_{0}$ for two identical atoms.

Generally speaking, for observing collectively enhanced atom-light coupling, an effective two-level  quantum system is used by neglecting the influence of the intermediate state with a large detuning \cite{Dudin12}. With such a simplified model, the transition to the doubly-excited state $|rr\rangle$ can be forbidden owing to a big energy shift induced by the interaction $U_{dd}$. 
This phenomenon is known as the excitation blockade \cite{Comparat10}, in which the frequency of collective Rabi oscillation between ground and singly-excited state is enhanced by a factor of $\sqrt{N}$ in the case of full blockade ($N$ is the number of atoms), compared to the single-atom case \cite{Lukin01}. The blockade effect has been applied to measure the strength of interactions between two collectively-excited atoms \cite{Gaetan09,Urban09,Wilk10}. Observing blockade effect requires a precise control of the interatomic distance $R$, which can be realized by changing the incidence angle of two trapping lasers in experiments. To our knowledge, the strength of excitation blockade can be classified by a universal criterion, according to the ratio between on the vdWs interaction $U_{dd}$ and the effective Rabi frequency $\Omega_{eff}$.
When $U_{dd}\gg\Omega_{eff}$, a full blockade takes place, in which $P_{r}(t)$ coherently oscillates at an enhanced frequency $\sqrt{N}\Omega_{eff}$ and $P_{rr}(t)\approx 0$; on the contrary, when $U_{dd}\ll\Omega_{eff}$, there is no blockade effect and each atom behaves independently with a same Rabi frequency $\Omega_{eff}$. If $U_{dd}$ and $\Omega_{eff}$ are comparable, $P_{r}(t)$ and $P_{rr}(t)$ behave more complexly, being sensitive to the exact value of $U_{dd}$. This criterion has been well-accepted and regarded as an essential condition for observing coherent many-body Rabi oscillation in a blockaded atomic ensemble.

In Section 3.1, with an effective two-level model (the large intermediate-detuning case) we propose a new way for determining the blockade strength, decided by comparing the numbers and values of Rabi oscillation frequencies. We show that the dynamical behaviors of Rabi oscillations in different blockade regions have similar plots with the experimental results as in \cite{Beguin13}. Moreover, in Section 3.2 we study the ultrafast two-photon Rabi oscillations for a small intermediate detuning on a timescale below 10$n$s by using more intense lasers, in order to overcome the strong spontaneous decay of $|m\rangle$. We present that the obtained Rabi cycles can be accelerated by three orders of magnitude ($\mu$s$\to$$n$s), offering more prospects for ultrafast population transfer and state manipulation with Rydberg atoms.

\subsection{Large intermediate-detuning case}  \label{onephotonex}

\begin{figure}
\centering
\includegraphics[width=3.5in,height=3.2in]{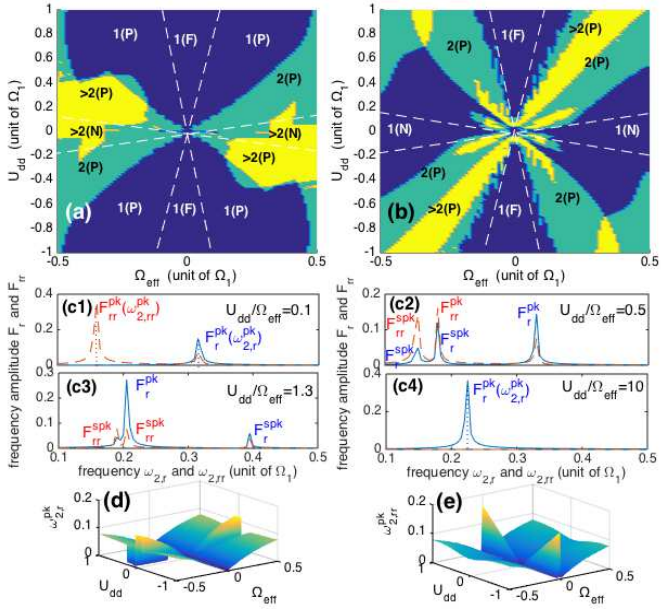}
\caption{(color online). Representation for the frequency numbers in Rabi oscillations between the ground state and (a) the singly-excited Rydberg state, (b) the doubly-excited Rydberg state, versus $\Omega_{eff}$ and $U_{dd}$. Labelings ``1", ``2", ``$>$2" denote the number of oscillating frequencies, and ``F", ``P", and ``N" denote the full, partial, and none blockade regimes. (c1-c4) The full frequency spectrum of $F_{r}(\omega_{2,r})$ (blue-solid) and $F_{rr}(\omega_{2,rr})$ (red-dashed) in the frequency domain for $U_{dd}/\Omega_{eff} = 0.1$ (none blockade), $0.5$ (partial blockade), $1.3$ (partial blockade) and $10$ (full blockade) where the amplitudes $F_{r(rr)}^{(s)pk}$ for the peak and sub-peak frequencies are given, and $\Omega_{eff}=1.0$MHz. (d-e) The corresponding peak frequencies $\omega_{2,r}^{pk}$ and $\omega_{2,rr}^{pk}$ with respect to (a) and (b). The parameter $\chi=1$ is used in all plots and $\Omega_{1}=20.0$MHz.}
\label{twoatom}
\end{figure}

\begin{figure}
\centering
\includegraphics[width=3.6in,height=2.3in]{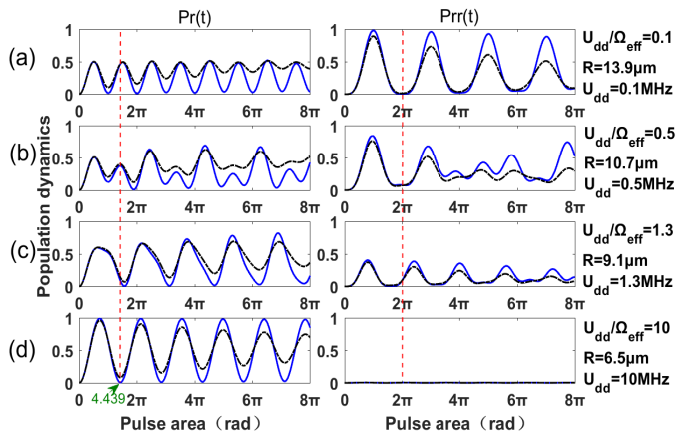}
\caption{(color online). (a-d) Time evolutions for the populations of the singly-excited state $P_{r}(t)$ (left panels) and the doubly-excited state $P_{rr}(t)$ (right panels). From top to bottom, as the ratio $U_{dd}/\Omega_{eff}$ increases ($\Omega_{eff}=1.0$MHz and vary $U_{dd}$), figures show the Rabi oscillations in the regimes of (a) none blockade, (b) partial blockade, (c) partial blockade , (d) full blockade, respectively. The cases of $\Gamma = 10$MHz (blue-solid) and $\Gamma = 200$MHz (black-dashed) are presented.}
\label{twoatomfre}
\end{figure}

The following numerical simulations of the atomic dynamics are performed on two $^{87}$Rb atoms with $|g\rangle=|5S_{1/2}\rangle$, $|m\rangle=|5P_{1/2}\rangle$, and Rydberg state $\left\vert r \right\rangle=\left\vert 62D_{3/2} \right\rangle$. The vdWs coefficient for $|62D_{3/2}\rangle$ is $C_{6}=2\pi\times 116.2 $GHz$.\mu$m$^6$ and the distance $R$ can be varied in a large range by controlling the directions of the lasers. 
For the large-intermediate-detuning case, $\left\vert m\right\rangle$ will experience a negligible population and the Rydberg state is excited by a two-level off-resonant Rabi oscillation between $\left\vert g\right\rangle$ and $\left\vert r\right\rangle$ with effective frequency $\Omega_{eff}$ and effective detuning $\delta_{eff}$. Besides, the amplitude of the Rabi oscillation suffers from an apparent damping due to the spontaneous decays while its oscillating frequency is unaffected. This point that we will explain it later in Fig. \ref{twoatomfre} by comparing the cases of $\Gamma=10$MHz (blue solid) and $\Gamma=200$MHz (black dashed).

The validity of such a two-level Rabi oscillation is guaranteed by an off-resonant excitation condition
\begin{equation}
\frac{\Omega_{eff}}{\Omega_{1}}\ll \frac{1}{2\chi\max[1,1/\chi]}\leq \frac{1}{2},
\end{equation}
deduced from the requirement of $\Delta\gg \Omega_1,\Omega_2$. We study the Rabi flopping under different $U_{dd}$s and $\Omega_{eff}$s, and record the numbers of oscillating frequencies in Fig. \ref{twoatom}(a-b), with ``F", ``P", ``N" in parentheses presenting full, partial and none blockade. The boundaries denoted by white dashed lines are extracted from a precise calculation of peak frequencies $\omega_{2,r}^{pk}$ and $\omega_{2,rr}^{pk}$, serving as a new criterion for classifying the blockade strength.
Specifically, the regime ``1(F)" is given by a single oscillating Rabi frequency with its precise value $\omega_{2,r}^{pk}= \sqrt{2}\Omega_{eff}$ and the regime ``1(N)" by a single peak oscillating frequency $\omega_{2,rr}^{pk}= \Omega_{eff}$. The ratio between them is expected to be $\sqrt{2}$ as same as the enhancement factor observed in ref. \cite{Beguin13}. In the middle regime of partial blockade labeled by ``2(P)" and ``$>$2(P)", $F_{r}$ and $F_{rr}$ are of multiple-peaked structures. To this end, see Fig. \ref{twoatom}(c2-c3), peak amplitudes $F_{r}^{pk}$, $F_{rr}^{pk}$ and sub-peak amplitudes $F_{r}^{spk}$,  $F_{rr}^{spk}$ are clearly shown. Fig. \ref{twoatom}(d-e) presents the distribution of peak frequency values $\omega_{2,r}^{pk}$ and $\omega_{2,rr}^{pk}$ in the space of $U_{dd}$ and $\Omega_{eff}$, revealing more elusive behaviors of the collective Rabi oscillations. In the partial blockade regime, the asymmetric and angular dependent interactions have been measured experimentally \cite{Barredo14}, arising collectively enhanced excitations that can even break the limit of $\sqrt{2}$ as for the full blockade regime \cite{Jing16}. For example, in (c3) the frequency $\omega_{r}^{spk}$ is almost two times as large as $\omega_{r}^{pk}$.

To verify the validity of this new classification we also plot complete frequency functions $F_{r}$ and $F_{rr}$ in the frequency domain in Fig. \ref{twoatom}(c1-c4) with tunable interactions $U_{dd}=0.1$MHz (none blockade), 0.5MHz(partial blockade), 1.3MHz(partial blockade), 10MHz(full blockade) and $\Omega_{eff}=1.0$MHz. Obviously, the oscillating frequency in (c1) and (c4) are dominated by single frequency $\omega_{2,rr}^{pk}$ and $\omega_{2,r}^{pk}$, respectively; while in (c2-c3) several comparable sub-peak frequencies $\omega_{2,r}^{spk}$ and $\omega_{2,rr}^{spk}$ arise, signifying more complex Rabi oscillations.


We further study the time dependence of population dynamics by varying the relative strength of $U_{dd}/\Omega_{eff}$ to meet the new criterion, and record a series of Rabi oscillations in the time domain in Fig. \ref{twoatomfre}, representing a comparative result with Fig. 2 in ref. \cite{Beguin13}. From top to bottom, we appropriately increase $U_{dd}$ via decreasing $R$ while keeping $\Omega_{eff}$ (=1.0MHz) unchanged, and the realistic parameters are displayed on the right. For two independent atoms [(a), none-blockade case], $P_{r}(t)$ is expected to oscillate between 0 and 1/2 at frequency $2\Omega_{eff}$, and $P_{rr}(t)$ oscillates between 0 and 1.0 at frequency $\Omega_{eff}$. In contrast [(d), full-blockade case], due to the strong vdWs interaction that induces a large energy shift to $\left\vert rr\right\rangle$, $P_{rr}(t)$ is substantially suppressed and at the same time $P_{r}(t)$ shows a collective Rabi oscillation with a perfectly enhanced frequency $(2\pi/4.439)\Omega_{eff}\approx\sqrt{2}\Omega_{eff}$ \cite{Gaetan09}. 
In addition, we compare the results by choosing $\Gamma=10$MHz (blue-solid) and 200MHz (black-dashed), and find that except for a damped amplitude with the increase of $\Gamma$, the oscillating frequency almost does not change at all. That is because a larger decay rate from the intermediate state will only give rise to a non-negligible decoherence for the population oscillation, companied by a quick damping to the oscillation amplitude, but the frequency is not influenced.

In the intermediate regime [(b-c), the partial-blockade cases], the imperfection of blockade gives rise to the fact that both $P_{r}(t)$ and $P_{rr}(t)$ show more complex and irregular Rabi oscillations. As a consequence, the exact frequencies of the oscillations in this regime are unpredictable and can even exceed the limit value $\sqrt{2}\Omega_{eff}$ of the full blockade case. This can be briefly understood with the aid of Fig. \ref{twoatom}(c3) where the sub-peak frequency $\omega_{2,r}^{spk}$ is found to be much larger than $\sqrt{2}\Omega_{eff}$ due to the excitation of doubly-excited Rydberg state. Finally, it is worthwhile to stress that our results (Fig.\ref{twoatomfre}) fully agree with the experiment data and theoretical analysis of Fig.2 in Ref. \cite{Beguin13} which were based on a pure two-level atomic system. This agreement strongly proves the significance of effective two-level model in the large intermediate-detuning case.

\subsection{Small intermediate-detuning case} \label{highfre}

In contrast to the effective Rabi oscillation, when the detuning $\Delta$ from the intermediate state is relatively small, the ground-Rydberg Rabi oscillation will suffer from a big damping, which is inimical to the study of collective excitation in Rydberg atoms. An efficient way to overcome this is utilizing an ultrashort or ultra-strong pulse of laser driving, enabling the population transfer to the target state on a very short timescale \cite{Avanaki16}. For this reason, the ground-Rydberg Rabi oscillation with a small intermediate detuning may become a shortcut to the generation of rapid quantum gate for quantum information process.

\begin{figure}
\centering
\includegraphics[width=3.6in,height=1.8in]{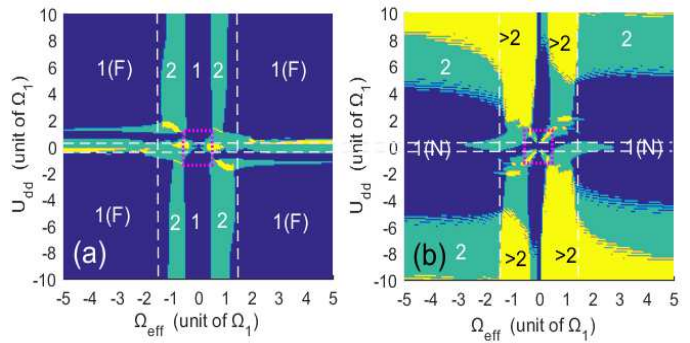}
\caption{(color online). Same as Fig.\ref{twoatom}(a) and (b) but extending to a wider parametric space of $\Omega_{eff}/\Omega_1\in[-5,5]$, $U_{dd}/\Omega_1\in[-10,10]$ and $\Omega_1=20$MHz. The parametric regions for Fig.\ref{twoatom}(a) and (b) are included in the red dotted box.}
\label{resnumb}
\end{figure}

For studying the collective Rabi frequency in a detailed way, we show the numbers of Rabi oscillating frequency in a wider parametric space in Fig. \ref{resnumb}(a-b), compared to the results of Fig. \ref{twoatom}(a-b) which are merely displayed in tiny dotted boxes.
In (a), $F_{r}(\omega_{2,r})$ is mostly a single-peak function as denoted by ``1(F)", dominated by one leading frequency $\omega_{2,r}^{pk}$ except in the regime where $U_{dd}/\Omega_{1}$ is very small, giving to the none blockade area, as denoted by ``1(N)" in (b). 
In contrast, comparing to $F_{r}(\omega_{2,r})$, $F_{rr}(\omega_{2,rr})$ [(b)] presents a single-peak structure only if $U_{dd}/\Omega_{1}$ is small. By increasing $|U_{dd}|$, $P_{rr}(t)$ will be deeply suppressed due to the full blockade effect, accompanying with complex oscillating frequencies and reduced amplitudes.

\begin{figure}
\centering
\includegraphics[width=3.4in,height=3.5in]{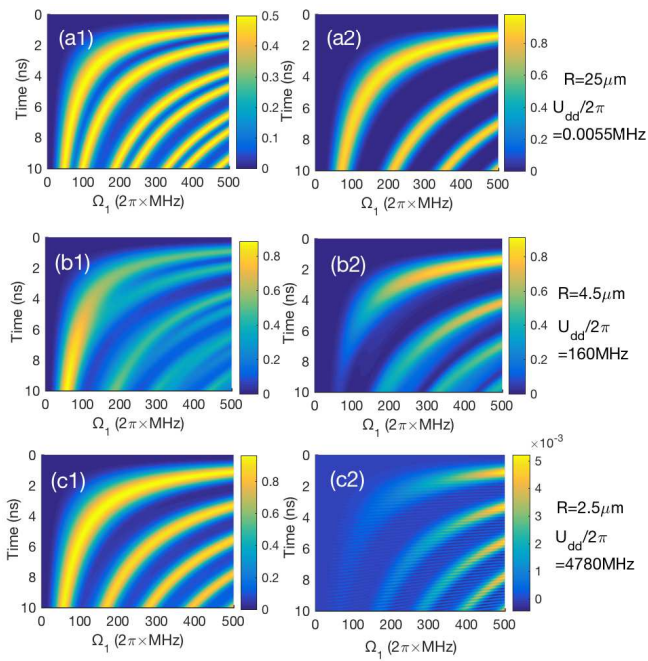}
\caption{(color online). Fast two-photon Rabi oscillations with the varying of the pump laser Rabi frequencies $\Omega_{1}$ and time $t$. Left panels: $P_{r}(t)$; Right panels: $P_{rr}(t)$. From (a) to (c) by tuning the interatomic distance $R=(25,4.5,2.5)\mu$m, the corresponding interaction strengths become $U_{dd}=2\pi\times(0.0055,160,4780)$MHz. Other parameters are described in the text.}
\label{fastosc}
\end{figure}

We further explore the fast dynamical behavior in a small-detuning Rabi oscillation, by performing a numerical simulation to the two-atom ME with experimental accessible parameters. For $^{87}$Rb atoms, here we adopt $\left\vert r \right\rangle=\left\vert 82D_{3/2} \right\rangle$, accordingly, the Rydberg decay $\gamma = 2\pi\times 5.75$kHz, the vdWs coefficient is $C_{6}=2\pi\times 1353.5($GHz$.\mu$m$^6$), the decay rate of the intermediate state $\left\vert 5P_{1/2} \right\rangle$ is $\Gamma=2\pi\times5.75$MHz. We assume,  both laser fields are resonant to their respective transitions with a same Rabi frequency, i.e. $\Omega_1=\Omega_2$. Differing from the off-resonant case, the atomic effective two-photon Rabi frequency is redefined by $\Omega_{res,eff}=\sqrt{\Omega_1^2+\Omega_2^2}/2=\Omega_1/\sqrt{2}$ for $\chi=1$.

Fast ground-Rydberg Rabi oscillations for $P_{r}(t)$ (left panels) and for $P_{rr}(t)$ (right panels) are presented in Fig. \ref{fastosc} with the variables $\Omega_1\in2\pi\times(0\sim500)$MHz and $t\in(0\sim10)$ns. From (a) to (c), different vdWs interactions are applied, that is, $U_{dd}=2\pi\times0.0055$MHz in (a1-a2), $U_{dd}=2\pi\times160$MHz in (b1-b2), and $U_{dd}=2\pi\times4780$MHz in (c1-c2). When $\Omega_1$ is close to $2\pi\times 500$MHz, we find that the frequency of the oscillation can reach to $\sim$GHz and the contrast to $\approx 95.4\%$.
For an intermediate state of a smaller linewidth, {\it e.g.} $6P_{1/2}$ with $\Gamma/2\pi = 1.3$MHz, this contrast can further be increased to $98.33\%$ as predicted by \cite{Bernien17}. Specifically, 
when $\Omega_1/2\pi=500$MHz (an intensity of $\sim2.0$MW/cm$^2$), for $U_{dd}=2\pi\times4780$MHz (c1-c2), the Rabi oscillating frequency for $P_{r}(t)$ is observed to be collectively enhanced to $\sqrt{3/2}\Omega_{res,eff}=2\pi\times 433$MHz. We find that, the enhancement factor $\sqrt{3/2}$ is slightly smaller than the full blockade enhancement factor $\sqrt{2}$, which is mainly caused by the quantum interference of two-photon optical transition paths. This fast Rabi oscillation enables an efficient population conversion to the Rydberg state within only a few $n$s. Besides, $P_{rr}(t)$ reveals a large suppression there, as a signature for that the system works in the full blockade region. Note that for same laser drivings, the exact value of $\sqrt{3/2}\Omega_{res,eff}$ is usually larger than $\sqrt{2}\Omega_{eff}$ according to their different definitions.

In the absence of interactions, for $U_{dd}=2\pi\times 0.0055$MHz (a1-a2) and still $\Omega_1/2\pi=500$MHz, $P_{r}$ and $P_{rr}$ oscillate with different frequencies and amplitudes, the ratio of frequencies between them is 2.0 and the ratio of amplitudes between them is 0.5, as same as expected in none blockade regime. The transitioning region (b1-b2) demonstrates compatible Rabi oscillations with several frequencies that depend on $U_{dd}$, as a signature for partial blockade. In this region we observe an imperfect population transfer with a rapid amplitude damping, which confirms the findings that the decays from the intermediate states can lead to the breakup of coherence in collective excitation.

Finally, we briefly analyze the dependence for the blockade strength in a two-photon collective excitation to Rydberg states. From Fig. \ref{fastosc}, we obtain that, for full blockade the single-valued Rabi oscillation frequency $F_{r}=\sqrt{3/2}\Omega_{res,eff}$ [(c1)] and for none blockade the single-valued frequency is $F_{rr}=\Omega_{res,eff}$ [(a2)], presenting the collective effect of two-atom excitation. Therefore, we re-stress that it is feasible to determine the numbers and values of Rabi oscillating frequency as a new criterion for classifying full and none blockade, as denoted by ``1(F)", ``1(N)"  in Fig. \ref{resnumb}.

\section{Conclusion}

We theoretically investigate the properties of collective Rabi oscillations of two three-level Rydberg atoms. Our results show that for a large-intermediate-detuning case, a straightforward collective excitation between the ground and the Rydberg state is established, but whether or not a high-contrast Rabi oscillation can be observed depends on the relative strength of the interatomic vdWs interaction $U_{dd}$ and the effective Rabi frequency $\Omega_{eff}$. The characteristic $\sqrt{2}$ scaling of the collective Rabi frequency enhancement is clearly verified, accompanied by a newly-proposed way to classify the blockade effect according to a detailed analysis for the numbers and values of the frequencies of Rabi oscillation. 
While turning to the small-intermediate-detuning regime, the collective effect of two-photon excitation is relative weak due to the increasing influence of the decay from the intermediate state. The resulting enhancement factor to the effective Rabi oscillation frequency is found to be smaller than $\sqrt{2}$. However, the reduction of the influence of the intermediate decay at the expense of high power of pumping lasers ($\sim$MW/cm$^2$) leads to a fast two-photon Rabi oscillation in an ultrashort timescale ($\sim n$s), opening possibilities to achieve ultrafast quantum state control and quantum logic gate generation.

Our theoretical results presented have an excellent match with the previous experimental and theoretical results, and can also provide detailed advices on the optimal parameters selection in the future experiments. The next-step work will be focused on simulations of ultrafast quantum logic gate and population transfer dynamics.

\acknowledgements

This work is supported by the NSFC under Grants No. 11474094, No. 11104076, and No. 11574086,
by the Science and Technology Commission of Shanghai Municipality under Grant No. 18ZR1412800 and No. 16QA1401600, 
the Specialized Research Fund for the Doctoral Program of Higher Education No. 20110076120004.

\appendix

\end{document}